\documentclass[preprint,preprintnumbers,nofootinbib,amsmath,amssymb,floatfix]{revtex4-2}

\usepackage[english]{babel}
\usepackage{color}
\usepackage{amsmath}
\usepackage{physics}
\usepackage{xr-hyper}
\usepackage[hidelinks]{hyperref}
\usepackage{diagbox}
\usepackage{multirow}
\usepackage{amsfonts}
\usepackage{amssymb}
\usepackage{latexsym}
\usepackage{graphicx}
\usepackage{adjustbox}
\usepackage{float}
\usepackage{xr}
\usepackage{subcaption}
\usepackage{caption}
\usepackage{enumitem}
\usepackage{threeparttable}
\usepackage{ytableau}
\usepackage[nodisplayskipstretch]{setspace} 
\newcommand{\al}{\alpha}
\newcommand{\be}{\beta}

\newcommand{\ga}{\gamma}

\newcommand{\de}{\delta}

\newcommand{\rar}{\rightarrow}
\newcommand{\lrar}{\leftrightarrow}

\newcommand{\non}{\nonumber}

\allowdisplaybreaks[3]

\begin{document}

\title{Ultra-Compact accurate wave functions for He-like iso-electronic sequences and variational calculus. IV. Spin-singlet states $(1s\,ns)$ $n\,{}^1 S$ family of the Helium sequence}

\author{J.C.~Lopez Vieyra}
\email{vieyra@nucleares.unam.mx}

\author{A.V.~Turbiner}
\email{turbiner@nucleares.unam.mx, alexander.turbiner@stonybrook.edu (corresponding author)}

\affiliation{Instituto de Ciencias Nucleares, Universidad Nacional Aut\'onoma de M\'exico,
A. Postal 70-543 C. P. 04510, CDMX, M\'exico.}


\setstretch{1.4}

\begin{abstract}
As a continuation of Parts I \cite{Part-1:2020}, II \cite{Part-2:2021}, III \cite{Part-3:2022}, where ultra-compact wave functions were constructed for a few low-lying states of He-like and Li-like sequences, the family of spin-singlet $(1s\,ns)$ type excited states $n\,{}^1 S$ of the He-like sequence is studied with an emphasis on the $n=3,4,5$: $3\,{}^1 S, 4\,{}^1 S, 5\,{}^1 S$ states, for nuclear charges $Z \leq 20$. Particular attention is given to finding of critical charges $Z=Z_B$ at which the ultra-compact wave functions lose their square-integrability.

For each ${}^1 S$ state an ultra-compact, seven-parametric trial function is constructed, which describes the domain of applicability of the non-relativistic Quantum Mechanics of Coulomb Charges (QMCC) for the total energies (4-5 significant digits (s.d.)) and reproduces
3 decimal digits (d.d.) of the spin-singlet states $n\,{}^1 S$ of He-like ions (in the static approximation with point-like, infinitely heavy nuclei) for $n=1,2,3,\ldots$ and any $Z \leq 20$\,. All energies are well described by second degree polynomials in $Z$ (the Majorana formula).

Critical charges $Z=Z_B^{(n)}$, where the ultra-compact trial function for the
$n^1 S, n=1,2,3,\ldots$  states loses its square-integrability, are estimated: for all studied states $Z_B^{(n)}$ increases slowly with $n$; it seems they lie in the interval $Z_B(n^1 S) \sim 0.90 - 0.95$, in particular, with $Z_B^{(1)}=Z_B^{(2)}\,=\,0.904$, $Z_B^{(3)}=Z_B^{(4)}\,=\,0.928$, $Z_B^{(5)}\ =\ 0.939$.
\end{abstract}

\enlargethispage*{10pt}

\maketitle

\newpage

\setstretch{1.5}

\section*{Introduction}

Many years ago Frank Harris (Gainsville-Utah) attracted our attention to the idea of constructing few-parametric, easily manageable numerically, highly-compact (ultra-compact) trial functions for the Helium sequence, which can provide sufficiently high accuracy in the total energies for any relevant physically, nuclear charge $Z$ \cite{Harris:2005}.
Recently, this idea was extended to obtain not only high accuracy in the total energies but also in the electron-nucleus and electron-electron cusp parameters (the Kato parameters) \cite{Part-1:2020}.
This idea can be modified by specifying the accuracy in energies: we want to describe the domain of applicability of the non-relativistic Quantum Mechanics of Coulomb Charges (QMCC), which is free of mass, relativistic and QED corrections, for the total energies of the Helium sequence with a minimal number of free parameters. This is the goal of the present article. The number of needed free parameters turned out to be {\it seven} for the spin-singlet $S$ states at $2 \leq Z \leq 20$:
this provides three correct decimal digits for the energy.

As a continuation of Part I \cite{Part-1:2020}, dedicated to the ground state of He-like and Li-like isoelectronic sequences for nuclear charges $Z \leq 20$, of Part II \cite{Part-2:2021}, dedicated to the spin-singlet excited state $2^1 S$ and the lowest spin-triplet $1^3 S$ of the He-like sequence, and of Part III \cite{Part-3:2022}, where two ultra-compact wave functions in the form of generalized Guevara-Harris-Turbiner functions were constructed for the spin-quartet state $(1s2s3s)$ of the Li-like sequence $1^40^+$, in the present paper the sequence of spin-singlet $(1s\,ns)$ excited states $n\,{}^1 S$ of the He-like sequence is studied. The ultra-compact wave functions for $n=1,2$ are revisited while those for the $n=3,4,5$: $3\,{}^1 S$\,,\, $4\,{}^1 S$ and $5\,{}^1 S$ states, for nuclear charges $Z \leq 20$ are presented in details. All constructed ultra-compact functions are built to reproduce the exact solutions in the limit of large $Z$. Particular attention is given to searching for critical charges $Z=Z_B$ where the ultra-compact wave function becomes non-normalizable. The Majorana formula (see below Ref.\cite{E-S:2012}) for the energy {\it versus} $Z$ is checked and continues to hold for the spin-singlet excited $S$ states other than those explored in Parts I-II.

It is worth noting two general observations, which can be deduced (and then conjectured to be true) from the studies in Parts I-III. If for finite nuclear charge $Z$ the accuracy in the total energies is limited to 3-4 significant figures, then:
(I) the pre-factor in the $n$th Coulomb orbital $(ns)$ (which is the exact eigenfunction of the two-body Coulomb problem), given by the $n$th associated Laguerre polynomial $L^{(1)}_{n-1}(2Zr/n)$, remains a $(n-1)$th degree polynomial $\sim P_{n-1}(r)$ and
(II) each Coulomb exponent $\sim r$ is replaced by a simple rational function $r Q_1(r)/{\tilde Q}_1(r)$, which is the ratio of a 2nd degree polynomial to a 1st degree polynomial in the variable $r$, this replacement models a distance-depended screening which corresponds to the two-body Coulomb interaction at small and large $r$.

\bigskip\bigskip

The Hamiltonian which describes the non-relativistic (two-electron) Helium-like atomic system with an infinitely-massive nucleus of charge $Z$ is given by
\begin{equation}
\label{H-He}
  {\cal H}\ =\ -\frac{1}{2} \sum_{i=1}^{2} \boldsymbol{\nabla}_i^2 \ -\ \frac{Z}{r_1} - \frac{Z}{r_2} \ +\ \frac{1}{r_{12}}  \ ,
  \end{equation}
where $\boldsymbol{\nabla}_{i}$ represents the 3-dimensional momentum of the
$i$-th electron (of mass $m=1$) situated at the position ${\mathbf r}_i$, $\hbar=1$;
the potential contains the attractive Coulomb interaction between the nucleus of charge $Z$ and each electron, and the Coulomb repulsion between the electrons. Here $r_i = |{\mathbf r}_i|$ is the distance from the nucleus to the $i$-th electron and $r_{12}=|{\mathbf r}_1 - {\mathbf r}_2|$ is the inter-electronic distance.

For concrete calculations of variational energies we used the direct numerical evaluation
of the six-dimensional integrals, by means of a numerical adaptive multidimensional integration routine (Cubature). Dynamical partitioning of the integration domain was employed, for a description see \cite{PR:2006}. As for the minimization we employed the routine MINUIT from CERN-LIB. A dedicated code was written in C++ with MPI paralellization. The program was executed in the cluster KAREN at ICN-UNAM (Mexico) using 140 Xeon processors with working frequencies of 2.70 GHz each. Each integration took a few seconds of wall time, but due to the flatness of the minimum, in seven-dimensional parameter space, the process of minimization was repeated numerously. Eventually, for any given state and chosen value of nuclear charge $Z$, it took a few hours of wall time to reach the minimum.

This paper is written in a self-contained form with
the following structure. In Section I, we construct the ultra-compact wave function for the Helium-like sequence for the spin-singlet state $3{}^1S$ with a pre-factor in the form of a second degree polynomial in $r$. The exact solution as $Z \rar \infty$ of the form $(1s3s) + (3s1s)$ is presented explicitly. The $1{}^1S$ and $2{}^1S$ states are revisited.
We analyze the square integrability of this ultra-compact function and find the corresponding critical charge $Z_B^{(3S)}$. The Majorana polynomial for the $3{}^1S$ state is found explicitly.

\bigskip\bigskip\bigskip\bigskip

In Section II we design the ultra-compact wave function for the spin-singlet state $4{}^1S$
for different $Z$ with a pre-factor in the form of a third degree polynomial in $r$. The exact solution as $Z \rar \infty$ of the form $(1s4s)+(4s1s)$ is described.
The square integrability of the ultra-compact function for the state $4{}^1S$ is analysed
and the critical charge  $Z_B^{(4S)}$ is found. The validity of the Majorana formula for the $4{}^1S$ state is checked. A similar analysis is made in Section III for the spin-singlet state $5{}^1S$. Each Section I-III is written in a self-contained form.

In Section IV a general ultra-compact function is proposed for the spin-singlet $n{}^1S$ state in the form $(1s\,ns')+(ns'\,1s)$ for arbitrary $Z > 1$  with a pre-factor in the form of the $(n-1)$th degree polynomial in $r$, which reduces to the exact solution at $Z \rar \infty$.
This function is multiplied by the $r_{12}$-dependent Jastrow factor in the exponential form (see Part II \cite{Part-2:2021}, Section 2.2 for a detailed description). The collection of critical charges for different $n$ is presented in Table. A hint for the analytic structure of the energy in the complex $Z$-plane near the level crossings and critical charges is given. Validity of the Majorana formula for different $n$ is checked.

Atomic units are used throughout this article.

\section{Spin-singlet  $ 3{}^1S $ state}

\subsection{Generalities and results}

Following the experience collected in Parts I-III \cite{Part-1:2020} - \cite{Part-3:2022},
the trial function for the spin-singlet  $3^1S$ state of the form $$(3s'\,1s) + (1s\,3s')$$
for a Helium-like atom with nuclear charge $Z$ is proposed like
\begin{equation}
\label{He3SPsi}
\resizebox{0.95\hsize}{!}{$\Psi_{3^1S}\ =\
\frac{1}{2}(1 + P_{12})
\left[ (1 - a_1  Z r_1 + a_2 Z^ 2 r_1^2 - a_3 Z r_2 + a_4 Z^2 r_1 r_2  +  b r_{12})\,
e^{-\al Z {\hat r}_1 - \beta Z {\hat r}_2  + \gamma {\hat r}_{12}}\right] \ ,$}
\end{equation}
where $P_{12}$ is permutation operator $(1 \lrar 2)$, $a_{1,2,3,4}, b, \al, \beta, \gamma$ are parameters. Note that this trial function (\ref{He3SPsi}) is a function of the relative distances $r_1,r_2,r_{12}$ only, no angular variables involved. Needless to say, that the function of this type can also be used to describe the $1{}^1S$ and $2{}^1S$ states, cf. Part I \cite{Part-1:2020} and Part II \cite{Part-2:2021}, respectively. In the limit $Z\rar \infty$, the function (\ref{He3SPsi}) reduces to the exact eigenfunction,
\[
  \Psi_{3^1S}^{(exact)}\ =\ (3s1s) + (1s3s)\ =\ \frac{1}{2}(1 + P_{12})\ \bigg(L^{(1)}_2(2Zr_1/3)
  \,e^{-\frac{Zr_1}{3} - Z r_2 }\bigg)\ ,
\]
where $L^{(1)}_2(r)$ is the second (associated) Laguerre polynomial with index 1, see below. It describes two non-interacting Hydrogen atoms: one is in $(1s)$ state and another one is in $(3s)$ state.

The effective distances ${\hat r}_{1,2}$ and ${\hat r}_{12}$, which appear in the exponential part of (\ref{He3SPsi}), are defined as rational functions,
\begin{equation}
\label{rationalrs}
  {\hat r}_1 = \frac{1+c_1 r_1}{1+ d_1 r_1}\ r_1\ , \quad {\hat r}_2 = \frac{1+c_2 r_2}{1+ d_2 r_2}\ r_2\ , \quad  {\hat r}_{12} = \frac{1+c_{12} r_{12}}{1+ d_{12} r_{12}}\ r_{12}\ ,
\end{equation}
where $c_{1,2,12}, d_{1,2,12}$ are parameters. This allows us to describe the screening of the Coulomb interaction between two charges at different distances which occurs due to the presence of the third charge. If $c=d$ this type of distant-depending screening is absent. At small $r$ the interaction is given by the interaction of bare Coulomb charges (no screening), while at large $r$ the screening is defined by the ratio $c/d$.

The coefficients $a_1, a_2$ in (\ref{He3SPsi}) are assumed to be determined by the orthogonality conditions: the function $\Psi_{3^1S}$ has to be orthogonal to both functions  $\Psi_{1^1S}$ and $\Psi_{2^1S}$,
\begin{align}
 \langle \Psi_{3^1S} | \Psi_{1^1S} \rangle &=0\ ,
 \non \\
 \langle \Psi_{3^1S} | \Psi_{2^1S} \rangle &=0\ ,
\label{ortho3S}
\end{align}
both of them, supposedly, are found beforehand \footnote{In principle, the orthogonality can be approximate, of the order of the relative accuracy, which we want to reach in the energies.}.
The concrete trial functions for the states $1^1S$ and $2^1S$ used in the orthogonality conditions (\ref{ortho3S}) will be described below. The remaining coefficients $a_3, a_4, b, \al, \beta, \gamma, c_1, d_1, c_2, d_2, c_{12}, d_{12}$ in (\ref{He3SPsi}) are treated as free, variational parameters.
Preliminary variational calculations indicate that the replacement $\hat {r_2} \to r_2$ in (\ref{He3SPsi}) does not change the variational energy in the first 3-4 significant digits (s.d.) for all $Z > 1$. Thus, we choose $c_2=d_2$ in all further calculations. It reduces the number of variational parameters by two: the total number of free parameters is equal now to ten. It is also true for the parameters $a_3, a_4, b$: they can be placed equal to zero without affecting the first 4-5 s.d.
 in the energies. The latter will be checked later, see Table I.
Eventually, the function (\ref{He3SPsi}) depends on seven free parameters, which will be defined variationally. Its final form reads
\begin{equation}
\label{He3SPsi-final}
\Psi_{3^1S}\ =\
\frac{1}{2}(1 + P_{12})
\left[ (1 - a_1  Z r_1 + a_2 Z^ 2 r_1^2)\,
e^{-\al Z {\hat r}_1 - \beta Z {r}_2  + \gamma {\hat r}_{12}}\right] \ .
\end{equation}
Note that asymptotically, at $Z \to \infty$ the exact wave function is given by
\begin{equation}
\label{3Sexact}
  \Psi_{3^1S}^ {\rm exact} = \frac{1}{2}(1 + P_{12}) \left(1 -\frac{2}{3} Z  r_1 + \frac{2}{27} Z^2 r_1^2 \right)\,
  e^{- \frac{Z}{3} {r}_1 -  Z r_2  }\ ,
\end{equation}
c.f. (\ref{He3SPsi-final}), with the energy
\footnote{It is more than well-known that for the Hydrogen atom with a nucleus of charge $Z$ the energy levels are given by ${\displaystyle E_{n}=-{\frac {Z^{2}}{2n^{2}}}}\ ,\ \hbox{n=1,2,\ldots} $ in a.u., where $n$ is the principal quantum number.
Thus, for a two electron atom in spin-singlet $n^1S$ state, where inter-electron interaction is neglected, the total energy is equal to
$E_1+E_n = - {Z^2}\left(\frac{1}{2} + \frac{1}{2n^2} \right)
= -\frac{5}{8}  Z^2, -\frac{5}{9}  Z^2, - \frac{17}{32} Z^2, -\frac{13}{25}  Z^2$
for $n=2,3,4,5$, respectively.}
\begin{equation}
E_{3^1S} = -\frac{5}{9} Z^2 \ ,
\end{equation}
which corresponds to two Hydrogen atoms: one is in its ground state while the another one is
in the excited state with principal quantum number $n=3$.
Hence, at $Z \to \infty$, the coefficients in (\ref{He3SPsi}) should behave in the following way
\begin{equation}
\label{3Sparams-asympt}
  a_1^{3S} \to \frac{2}{3}\ ,\ a_2^{3S} \to \frac{2}{27}\ ,\ a_{3,4}^{3S}\to 0\ ,
\ b^{3S}\to 0\ ,
\end{equation}
\[
\alpha^{3S}\to 1/3\ ,\ \beta^{3S}\to 1\ ,\ \gamma^{3S}\to 0\ .
\]
Note that the approach of $\gamma^{3S}$ to zero appears in a very slow pace, even
at $Z=20$ this parameter is still not small, see below.
Additionally, since ${\hat r}_1\to r_1$,   ${\hat r}_2\to r_2$, it is implied
\[
c_1=d_1\ ,\ c_2=d_2\ .
\]


The variational energies for the ${3^1S}$ state for $Z=2, 5, 10, 15, 20$ (taken as examples) based on the trial function (\ref{He3SPsi}) are
presented in Table~\ref{Table1} for two different configurations: (i) with $a_3=a_4=0$, $b=0$  (first row)
\footnote{Variational calculations with $b \neq 0$ showed that the variational parameter
$b$ is rather small and the energy improvement appears in the 4th-5th s.d. It can be placed safely equal to zero.},
and (ii) with  $a_3, a_4$ determined variationally and again $b=0$ (second row).
For comparison, we also include in  Table~\ref{Table1} results of calculations from  \pagebreak
(a) Accad, Pekeris and Schiff (APS) \cite{APS:1971} (which were done with an expansion
of the wave function in a triple series of the Laguerre polynomials of the perimetric coordinates with
220-364 terms included), (b) from Drake and Yan (DY) \cite{Drake:1994,Drake:2006} obtained with a double basis set
in Hylleraas coordinates with up to $\sim$ 1000 terms included
\footnote{ For $Z=2$ the non-relativistic energy for the ${3^1S}$ state calculated by DY is
  $E= -2.061\, 271\, 989\, 740\, 893\, 0$ with $N_{\rm max} = 937$ terms and
  $E= -2.061\, 271\, 989\, 740\, 911\, 3(5)$ (extrapolated).},
and (c) from  Aznabaev, 
Bekbaev and Korobov (ABK) \cite{Korobov:2018} $\sim 10^4$ terms in the exponential expansion included
(the benchmark results)
\footnote{ For $Z=2$ the non-relativistic energy for the ${3^1S}$ state calculated by ABK
  is reported with two different numbers of terms in the so-called exponential expansion:
  (a) with \hbox{$N=18000$}, $E=-2.061\,  271\,  989\,  740\,  908\,  650\,  740\,
  349\,  937\,  089\,  281\,  6 $ and,
  (b) with \hbox{$N=22000$}, \hbox{$E=-2.061\,  271\,  989\,  740\,  908\,  650\,  740\,  349\,  937\,  089\,  282\,  4 $}. It indicates the
  correctness of 34 s.d. Such an accuracy is much beyond of the existing experimental data and the theory at present times.}.

\bigskip

It should be noted that at $Z=2$ by making the comparison of the results available in literature with ABK results, which are benchmark at the moment, one can see that the accuracy of APS and DY results is grossly exaggerated: the 6th s.d. in APS energy is found wrongly (and all subsequent digits) and in DY energies the 14th s.d. is incorrect (and
all subsequent digits). It must be emphasized that the relativistic and QED corrections to the total energy can change 4th-5th s.d., see \cite{AOP:2019}. Eventually, we will employ the ultra-compact function (\ref{He3SPsi-final}) with 7 free parameters
$\al, \beta, \gamma, c_1, d_1, c_{12}, d_{12}$. It must be emphasized that the energies obtained with (\ref{He3SPsi-final}) differ from accurate (exact) values in 4-5 d.d., see Table~\ref{Table1}, similarly to what happened for ${1^1S}$ and ${2^1S}$ states, see below.
The energies at $Z>10$ for $3^1S$ state are found for the first time. We assume that they
are accurate in 3 d.d. at least.

\begin{table}[htp]
\caption{\label{Table1}
         {\it Variational energy $E$ (a.u.) of the $3{}^1S$ state of He-like
              isoelectronic sequence at $Z=3/2,2,5,10,15,20$ obtained with trial function
              (\ref{He3SPsi}) with optimized parameters when $b=0, c_2=d_2$ in two configurations:
              when $a_3=a_4=0$ (first row, function (\ref{He3SPsi-final})) or when
              $a_3, a_4$ are optimal (second row).
              Comparison of the our results with ones by APS \cite{APS:1971} rounded to 6 s.d.
              and marked by ${}^\dagger$ at $Z=2,10$, with ones by DY  \cite{Drake:1994,Drake:2006} marked by ${}^\ddagger$, and by \cite{Korobov:2018}(rounded) marked by ${}^\S$  at $Z=2$ presented}}
\begin{center}
{\setlength{\tabcolsep}{0.1cm} \renewcommand{\arraystretch}{1.2}
 \resizebox{0.9\textwidth}{!}{%
\begin{tabular}{|l|l|l|l|l|l|l|} \hline
 $Z$        &\ \ ${3}/{2}$  &\ \ $2$ &\ \ $5$ &\ \ $10$ &\ \ $15$ &\ \ $20$  \\
\hline\hline
$E$  &\ -1.14155 &\ -2.061 20                           &\ -13.41169 &\ -54.55195          & \ -123.47004 &\ -220.16593  \\
                 & &\ -2.061 227                          &            &\ -54.55203           &            &\ -220.1660  \\
           &&\ -2.061 264${}^\dagger$              &            &\ -54.55212${}^\dagger$ &            &  \\
           &&\ -2.061 271 989 740 893 ${}^\ddagger$ &            &                       &            &  \\
           &&\ -2.061 271 989 740 909 ${}^\S$       &            &                       &            &
\\[5pt]
$a_1 $    &0.56293&  0.45493                            &  0.585298  & 0.626149              &  0.639798   &  0.646526 \\
               &&  0.438952545                    &                   & 0.623057578        &                    &  0.6452137375
\\[5pt]
$a_2$      &0.04297&  0.0302925                         & 0.054744  & 0.0640833              & 0.0673567 & 0.069015275   \\
                &&  0.0290066                         &                  & 0.0637954              &                   & 0.068893605
\\[5pt]
$a_3$      &\ \ 0   & \ \ 0                                      &  0             &   0                             &  0             &  0                    \\
                &&  0.01745079                        &                  &  0.00293301            &                 &  0.001018747
\\[5pt]
$a_4$  &\ \ 0   & \ \ 0                                          &   0             &    0                            &   0            &  0 \\
           &&  0.00154585                             &                   &   0.000199878          &                 & 0.000012495
\\[5pt]
$\alpha$   &0.71596&  0.60513                             &   0.43291  & 0.37537               &  0.35853   & 0.34962   \\
           & & 0.53185516                          &            & 0.35564192            &            & 0.34649846
\\[5pt]
$c_1$      &0.10049&  0.21043                             &  0.83838   & 1.92958               &  2.75929   & 3.24933   \\
           &&  0.23555307                          &            & 1.45953784            &            & 3.38773032
\\[5pt]
$d_1$      &0.56494&  0.71707                             &  1.35834   & 2.41340               &  3.18436   & 3.59643    \\
           &&  0.70013372                          &            & 1.73602008            &            & 3.71213233
\\[5pt]
$\beta$    &0.9997&  0.99929                             &  0.9994    & 0.99943               &  0.99964   & 0.99970  \\
           &&  1.00292756                          &            & 0.99989304            &            & 0.99978757
\\[5pt]
$\gamma$   &0.68353&  0.58276                             & 0.52408    & 0.48658               &  0.46826   & 0.45684 \\
           &&  0.58387657                          &            & 0.42564680            &            & 0.44351786
\\[5pt]
$c_{12}$   &-0.0012& -0.00486                             & -0.03392   & -0.09652              &  -0.16934  & -0.24587 \\
           && -0.00304945                          &            & -0.1103450            &            & -0.22306295
\\[5pt]
$d_{12}$   &0.62392& 0.63739                              & 1.10623    & 1.84348               &  2.48298   & 3.0180    \\
           && 0.711559816                          &            & 1.60389096            &            & 3.71213233
\\[5pt]
\hline
\end{tabular}}}
\end{center}
 \end{table}%



The results presented in Table~\ref{Table1} for the state $3{}^1S$ correspond to the calculations with the function $\Psi_{3^1S}$  (\ref{He3SPsi}) which is constructed in such a way to be orthogonal to $\Psi_{1^1S}$ and  $\Psi_{2^1S}$ states: these two functions are of the same functional form as (\ref{He3SPsi}). Now we can present the concrete parameters for
ultra-compact functions for the spin-singlet states $\Psi_{1^1S}, \Psi_{2^1S}$ and  $\Psi_{3^1S}$ which were used for calculations of the total energy in domain $Z \in [2, 20]$\footnote{For all three states there were carried out the detailed calculations
for $Z=1.4, 1.5, 2, 3, 4, 5, 10, 15, 20$}. For convenience, these parameters are fitted by some simple functions of $Z$. For practical reasons the behavior of the parameters for larger $Z > 20$ is not studied.

\clearpage

$\bullet$\ {\it ${1^1S}$ state revisited.}

For the state  $\Psi_{1^1S}$  (c.f. $\Psi_{\sf F}$, Eq. (23) in \cite{Part-1:2020}),  we take the function similar to (\ref{He3SPsi}),
\begin{equation}
\label{He1SPsi}
\resizebox{0.95\hsize}{!}{$\Psi_{1^1S}\ =\
\frac{1}{2}(1 + P_{12})
\left[ (1 - a_1  Z r_1 + a_2 Z^ 2 r_1^2 - a_3 Z r_2 + a_4 Z^2 r_1 r_2  +  b r_{12})\,
e^{-\al Z {r}_1 - \beta Z {r}_2  + \gamma {\hat r}_{12}}\right] \ ,$}
\end{equation}
where $P_{12}$ is permutation operator $(1 \lrar 2)$, $a_{1,2,3,4}, b, \al, \beta, \gamma$ are parameters. Here
\begin{equation*}
\label{rationalrs-1S}
  {\hat r}_{12} = \frac{1+c_{12} r_{12}}{1+ d_{12} r_{12}}\ r_{12}\ ,
\end{equation*}
where $c_{12}, d_{12}$ are parameters. Accurate variational calculations demonstrate that
the optimal (non-zero) parameters $a_{2,3,4}$ influence the 4th-5th (and further) d.d. in the energies: they can be vanished without affecting the first three decimal digits in the energies. It leads to the seven-parametric ultra-compact trial function
\begin{equation}
\label{He1SPsi-final}
\Psi_{1^1S}\ =\
\frac{1}{2}(1 + P_{12})
       \left[ (1 - a_1  Z r_1  +  b r_{12})\,
        e^{-\al Z {r}_1\ -\ \beta Z {r}_2\ +\ \gamma \, r_{12}\,
        \frac{1+(d_{12}-\delta_{12}) r_{12}}{1+ d_{12} r_{12}}}
       \right] \ ,
\end{equation}
that provides systematically the variational energies with 3 d.d. correctly,
see Table \ref{Table-GS}: the difference with established results occurs in the fourth d.d.

\begin{table}[htb]
\caption{\label{Table-GS}
          {\it Variational energies $E_{var}$ (a.u.) of the $1{}^1S$ state of He-like
          isoelectronic sequence obtained with trial function (\ref{He1SPsi-final})
          together with optimal parameters compared to the established energies
          $E^{\rm exact}$ (a.u., rounded to 6d.d.), collected in \cite{AOP:2019}.}}
 \begin{center}
 {\small
\begin{tabular}{|c|c||c||c|c|c|c|c|c|c|} \hline
 $Z$          &  $E^{\rm exact}$     &  $E_{var}$         &  $\al$ & $\beta$ & $\gamma$ &  $a_1 Z$ & $b$     & $d_{12}$  & $\delta_{12}$     \\ \hline\hline
1              &  $-0.527\, 751\ $   &  $-0.527\,23$     &  0.97248  & 0.44134 & -0.23549 & 0.27447 & 0.57121 & 0.13695  & 0.14037          \\
$\frac{3}{2}$  &              & $-1.464\,79 $  & 1.00749 & 0.61042 & -0.33236 & 0.29673 & 0.69579 & 0.18787   & 0.17623          \\
2             &  $-2.903\,724\ $ & $-2.903\,18 $  & 1.02051 & 0.68442 & -0.42355 & 0.32146 & 0.78899 & 0.21886 & 0.19967          \\
3             &  $-7.279\, 913\ $ & $-7.279\,31 $ & 1.03152 & 0.76042 & -0.59691 & 0.36301 & 0.97139 & 0.24945 & 0.23818          \\
4             &  $-13.655\, 566\ $ & $-13.654\,94 $ & 1.03626 & 0.80139 & -0.78868 & 0.39441 & 1.16557 & 0.33725 & 0.31140          \\
5             &  $-22.030\, 972\ $ & $-22.030\,33 $ & 1.03766 & 0.82706 & -0.95142 & 0.42522 & 1.33401 & 0.37221 & 0.35740          \\
10            &  $-93.906\, 807\ $ & $-93.906\,10 $ & 1.03944 & 0.88846 & -1.50023 & 0.46399 & 1.88443 & 0.55267 & 0.55267          \\
15            &  $-$               & $-215.781\,31$ & 1.03848 & 0.91222 & -1.84642 & 0.46990 & 2.22582 & 0.67194 & 0.67371          \\
20            &\ $-387.657\,234$ \ & $-387.656\,34$ \ & 1.03805 & 0.92538 & -2.05925 & 0.44595 & 2.43576 & 0.61994 & 0.72499          \\
\hline
\end{tabular}
                }
\end{center}
\end{table}

In domain $Z \in [1, 20]$ the variational parameters can be easily fitted (cf. Eqs.(27)-(28) in  \cite{Part-1:2020}):
\begin{align}
 \label{params1Sfits}
 \al_{{1S}}^{\rm fit}\, Z &\ =\ -0.03106  +  1.04049\,  Z\ ,  \non\\
 \beta_{{1S}}^{\rm fit}\, Z  &\ =\ -0.57141  +  0.95137 \, Z\ ,  \\
 \gamma_{{1S}}^{\rm fit}\, Z &\ =\ -5.34847  + 8.55565  \sqrt{Z} - 3.70018 \,Z\ ,  \non\\
 d_{12_{{1S}}}^{\rm fit} &\ =\ 0.06685  + 0.07227 \, Z - 0.00221\,  Z^2\ ,  \non\\
 \delta_{12_{{1S}}}^{\rm fit} &\ =\ 0.05710  + 0.07767\,Z - 0.00246\,  Z^2\ ,  \non\\
 a_{1_{{1S}}}^{\rm fit}\,Z &\ =\ 0.11202 + 0.18543\,{\sqrt{Z}} - 0.02369\,Z \ , \non\\
 b_{{1S}}^{\rm fit} &\ =\ 0.01551 + 0.54322\,{\sqrt{Z}} + 0.01000\,Z \ . \non
\end{align}
Note that the parameters $\al\,Z , \beta\,Z$ behave linearly on $Z$.
By taking fitted parameters (\ref{params1Sfits}) and plugging them into (\ref{He1SPsi-final}) may lead to a (slight) deterioration of the variational energies, c.f. Table \ref{Table-GS}.
The function (\ref{He1SPsi-final}) is the most accurate among seven-parametric so-far-known trial functions for the ground state of the Helium atom.


\medskip

$\bullet$\ {\it ${2^1S}$ state revisited.}

For the state  $\Psi_{2^1S}$  (c.f. $\Psi_{\sf G}$,  Eq. (12) in \cite{Part-2:2021}) we choose
\begin{equation}
\label{He1SPsi-2S}
\Psi_{2^1S}\ =\
   \frac{1}{2}(1 + P_{12})\,
   \left[ (1 - a_1  Z r_1)\,
    e^{-\al Z {\hat r}_1 - \beta Z {r}_2  + \gamma {\hat r}_{12}}\right] \ ,
\end{equation}
where $P_{12}$ is permutation operator $(1 \lrar 2)$, $a_1, \al, \beta, \gamma$ are parameters. Here
\begin{equation*}
\label{rationalrs-2S}
  {\hat r}_{1} = \frac{1+c_{1} r_{1}}{1+ d_{1} r_{1}}\ r_{1}\quad ,\quad
  {\hat r}_{12} = \frac{1+c_{12} r_{12}}{1+ d_{12} r_{12}}\ r_{12}\ ,
\end{equation*}
where $c_{1}, d_{1}, c_{12}, d_{12}$ are free parameters.
This function provides systematically the variational energies with 3 d.d. correctly,
see Table \ref{Table-2S}: the difference with established results occurs in the 4th decimal digit, similarly to $1^1S$ state, see Table \ref{Table-GS}.


\begin{table}[htb]
\caption{\label{Table-2S}
          Variational energy $E_{var}$ (a.u.) of the $2{}^1S$ state of He-like
          sequence obtained with trial function (\ref{He1SPsi-2S}) compared to established
          energies $E^{\rm exact}$ (a.u.) for $Z=2, 5, 10, 15, 20$.}
 \begin{center}
 {\setlength{\tabcolsep}{0.1cm} \renewcommand{\arraystretch}{1.5}
 \resizebox{1.0\textwidth}{!}{%
\begin{tabular}{|c|l||l||c|c|c|c|c|c|c|c|} \hline
 $Z$ &  $E^{\rm exact}$          &  $E_{var}$         &  $\al$ & $\beta$  & $\gamma$  &  $a_1 Z$   & $c_1$  & $\delta_1$&$c_{12}$ & $\delta_{12}$     \\
\hline\hline
$\frac{3}{2}$ &  & $-1.166\, 28$ & 0.99800 & 0.99898 & 0.69149 & 0.45512 & 0.19055 & 0.72500    & -0.00190 & 0.66838  \\
2      &  $-2.145\, 974\ $  ${}^{a}$ & $-2.145\,70 $ & 0.85372 & 0.99750 & 0.60532 & 0.71959          & 0.38297 & 0.78722 & -0.00622 & 0.72790  \\
5      &  $-14.578\,528\ $  ${}^{a}$ & $-14.577\,91 $ & 0.59992 & 0.99709 & 0.51554 & 2.23454      & 1.37504 & 0.64980 & -0.05707 & 1.22905  \\
10     &  $-60.295\,341\ $  ${}^{a}$ & $-60.294\,59 $ & 0.53571 & 0.99818 & 0.48362 & 4.74082    & 2.54027 & 0.47429 & -0.16341 & 2.07473  \\
15     &  $-137.261\,086\ $ ${}^{b}$ & $-137.260\,70 $ & 0.52039 & 0.99862 & 0.47456 & 7.24168
& 3.72244 & 0.42379 & -0.24544 & 3.06045  \\
20     &  $-245.476\,928\ $ ${}^{c}$ & $-245.476\,72$ & 0.51481 & 0.99902 & 0.46790 & 9.74290     & 4.75956 & 0.39966 & -0.38846 & 3.79518  \\
\hline
\end{tabular}}}
		\begin{tablenotes}
			\footnotesize
            \item ${}^*${Exact energies rounded to 6 d.d. taken from ${}^{a,b,c}$,
            in Eq.(\ref{He1SPsi-2S}) $d_{1}=c_{1}+\de_{1}$ and $d_{12}=c_{12}+\de_{12}$: \\
            ${}^{a}$  Yerokhin and Pachucki,  {\it Phys.Rev.\bf A 81} (2010)  \cite{Pachucki:2010} (for $Z=2$, $E= -2.145 974 046 054 417 415 799\,a.u.$), \newline
            ${}^{b}$  Drake,  {\it Springer Handbooks of Atomic, Molecular, and Optical Physics} (2006) \cite{Drake:2006}, \newline
             ${}^{c}$ Drake,  {\it Can.Journ.Phys.\bf 66} (1988) \cite{Drake:1988}
           }
		\end{tablenotes}
\end{center}
\end{table}

In domain $Z \in [2, 20]$ the variational parameters can be fitted like (c.f. Eq. (13) \footnote{This fit corresponds to constrained results when the cusp parameters are set to be exact, see \cite{Part-2:2021}.} in  \cite{Part-2:2021}):
\begin{align}
 \al_{{2S}}^{\rm fit} \, Z &=  0.761468 + 0.466883 Z\ ,  \non\\
 \beta_{{2S}}^{\rm fit}\, Z   &=  -0.002773 + 0.998626 Z\ , \label{params2Sfits}\\
 \gamma_{{2S}}^{\rm fit}\,Z &= 0.300996 + 0.454227 Z\ ,  \non\\
 c_{1_{{2S}}}^{\rm fit} &= - 0.58317 + 0.52068 Z - 0.01757  Z^2  \ ,  \non\\
 \delta_{1_{{2S}}}^{\rm fit} &= \,  0.796636 -0.011693 Z  -0.000971751 Z^2\ ,  \non\\
 c_{12_{{2S}}}^{\rm fit} &= 0.027317 - 0.016149 Z  -  0.000213 Z^2\ ,  \non\\
 \delta_{12_{{2S}}}^{\rm fit} &= 0.40869 +0.16318 Z  +  0.000637 Z^2\ , \non \\
 \non \\
  a_{{1_{2S}}}^{\rm fit} &=  -0.28621 + 0.50413 Z - 0.000142 Z^2  \non\ \ . \
\end{align}

It is important to emphasize that the parameters $ \al_{{2S}}^{\rm fit} \, Z,  \beta_{{2S}}^{\rm fit} \, Z,  \gamma_{{2S}}^{\rm fit} \, Z$ behave linearly with $Z$.
The parameter $a_1$, presented by fit in (\ref{params2Sfits}), is not variational, it is chosen to guarantee the orthogonality of $\Psi_{2^1S}$ and  $\Psi_{1^1S}$,
\begin{equation}
\label{ortho1s2s}
\langle \Psi_{2^1S} | \Psi_{1^1S} \rangle = 0\ .
\end{equation}
Note that taking into account seven fitted parameters in (\ref{params2Sfits}) for the $2{}^1S$ state and seven fitted parameters (\ref{params1Sfits}) for the state $1{}^1S$,  the relations (\ref{ortho3S}) allow us to determine the parameters $a_1^{3S}, a_2^{3S}$ for the $3^1S$ state, see (\ref{He3SPsi-final}). In domain $Z \in [2, 20]$ the fit of variational parameters for the $3^1S$ state trial function (\ref{He3SPsi-final}) read:
\begin{align}
\alpha_{3S}^{\rm fit}\,Z & = 0.570384 + 0.319942 Z\ , \non\\
\beta_{3S}^{\rm fit}\,Z  & =  -0.000683 + 0.999635 Z  \ ,
\label{params3Sfits}\\
\gamma_{3S}^{\rm fit}\,Z & =  0.246265 + 0.459639 Z\ ,
\non\\
 c_{1_{3S}}^{\rm fit} & = -0.348464 + 0.286014 Z - 0.004971 Z^2\ ,
\non\\
 d_{1_{3S}}^{\rm fit} & =0.119389 + 0.302642 Z - 0.006139 Z^2\ ,
\non\\
 c_{{12}_{3S}}^{\rm fit} & = 0.015005 - 0.010071 Z - 0.000086 Z^2\ ,
\non\\
  d_{{12}_{3S}}^{\rm fit} & = 0.363961 + 0.144783 Z + 0.000464 Z^2\ ,
\non \\
\non \\
  a_{1_{3S}}^{\rm fit} & =-0.437281 + 0.672023 Z - 0.000198 Z^2\ ,
\non\\
  a_{2_{3S}}^{\rm fit} & = 0.029066 - 0.102043 Z + 0.074002 Z^2\ .
\non
\end{align}
Needless to say that these fits should not be used for $Z>20$, since they require modifications in order to correspond to the correct asymptotic behavior at $Z \rar \infty$, see (\ref{3Sexact}) and (\ref{3Sparams-asympt}).

\subsection{Majorana formula for $3^1S$ state}

Many years ago Ettore Majorana (unpublished, for discussion see \cite{E-S:2012})
conjectured that a second degree polynomial in $Z$ can approximate 3-4 significant figures
in the ground state energy of the helium-like system versus the nuclear
charge $Z$. We called it the {\it Majorana} formula \cite{AOP:2019}. This was successfully
checked for the ground state of the helium and lithium sequences in \cite{AOP:2019}.
In Parts I-III [1-3] it was shown that the Majorana formula holds not only for the ground state
but also for some excited states including $2^1S$ state.
Using the variational energies calculated with trial function (\ref{He3SPsi-final}) with fitted parameters (\ref{params3Sfits}) presented in Table \ref{Emajo2}, one can show that the Majorana formula
of the form,
\begin{equation}
\label{Majorana3S}
   E(Z)\ =\ -\frac{5 Z^2}{9}\ +\ 0.105306 Z-0.0495003\ ,
\end{equation}
holds for the energy of the $3^1S$ excited state as well, see Table \ref{Emajo2}. Furthermore,
the formula (\ref{Majorana3S}) allows to reproduce 3-4 d.d. in the total energies for $1.5 \leq Z \leq 20$.

\begin{table}
\caption{\label{Emajo2} Variational energies $E_{var} \equiv E_{3^1S}$ (a.u.) of the $3^1S$ state calculated for different $Z$ based on the function (\ref{He3SPsi-final}) with parameters (\ref{params3Sfits}) {\it
compared} with ones of the Majorana formula  (\ref{Majorana3S}).   }
\begin{center}
\begin{tabular}{ccc} \hline
 $Z$ & $E_{var}$     & $E_{Majorana}$ \\
\hline\hline
 2   &   -2.06112     &   -2.06111   \\
 3   &   ---          &   -4.73358   \\
 4   &   ----         &   -8.51717   \\
 5   &  -13.41168     &  -13.41186   \\
 10  &  -54.55195     &  -54.55200   \\
 15  & -123.47004     & -123.46991   \\
 20  & -220.16593     & -220.16560   \\
\hline
\end{tabular}
\end{center}
\end{table}

\subsection{Square-integrability of the $3{}^1S$ excited state}

It is evident that the normalizability/square-integrability of the wave function (\ref{He3SPsi}) is guaranteed if it decays exponentially at large distances $r_1,r_2$.
In particular, this condition has to be valid whenever the position of one of the electrons is kept fixed  while the position of the other one tends to infinity. It can be explicitly seen in analysis of the exponential factors that
\begin{equation}
\label{Adef}
  \Psi_{{3}^1S} \big\vert_{r_i \ \rm fixed}  \to {\rm exp}\{- A_j  r_j \} \ \ \text{at}
  \ \ r_j\to\infty , \quad  i,j=1,2 \quad
(i\neq j)\ ,
\end{equation}where the factors $A_{1,2}$ depend on $Z$, they are positive $A_j > 0$, for $Z > Z_B$,
where $Z_B$ is called the {\it the second critical charge} for the $3{}^1S$ excited
state.  For the trial function (\ref{He3SPsi}) \footnote{In the limit when both $c_{i}, d_{i}$ tend to zero ($i=1,2,12$) the ratio $\frac{c_i}{d_i}=1$ must be taken.}
\begin{align}
\label{Asdefinition}
 A_1 & = \alpha Z \frac{c_1}{d_1} - \gamma \frac{c_{12}}{d_{12}} \equiv A_1^{3S} \ , \\
 A_2 & = \beta  Z \frac{c_2}{d_2} - \gamma \frac{c_{12}}{d_{12}} \equiv A^{3S}_2\ .\non
\end{align}
These quantities $A_1, A_2$ serve as a measure of the rate of convergence of involved integrals defining the square-integrability of $\Psi_{3{}^1S}$. The
trial function $\Psi_{3{}^1S}$  (\ref{He3SPsi}) remains normalizable as long as both
$A_1,A_2$ are positive. Thus, a critical charge $Z=Z_B$ is defined by the instance when either $A_1$ or $A_2$ vanishes first. The inverse of $A_j$ has a meaning of the average distance of the $j$-th electron to the nucleus, $\propto 1/A_j$. Our goal is to find such a critical charge $Z_B$.

\begin{figure}[hbt]
    \includegraphics[width=0.5\textwidth ,angle=0]{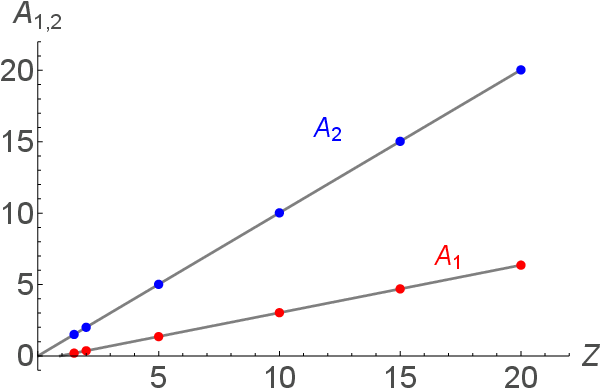}
\caption{\label{A1A2f3S}
    Parameters $A_1^{3S}, A_2^{3S}$ for different $Z$ and their interpolations (\ref{A1-3S})-(\ref{A2-3S})}
\end{figure}

By taking the variational parameters in $\Psi_{3^1S}$ (\ref{He3SPsi-final}) found for different $Z$ one can construct the plots of $A_{1,2}$ {\it vs.} $Z$, see Fig. \ref{A1A2f3S}. In general, both $A_1,A_2$ behave as linear functions in $Z$ with a slight deviation from linearity at small $Z$. They can be easily interpolated:
\begin{equation}
\label{A1-3S}
  A_1^{3S} = \frac{\alpha^{3S}  c_1^{3S} Z}{d_1^{3S}}-\frac{\gamma ^{3S} c_{12}^{3S}}{d_{12}^{3S}}\
  =\ \frac{1}{3} \left(Z - 0.928\right) \ -\ 0.000109\ \sqrt{Z-0.928}\ ,
  \end{equation}
\begin{equation}
\label{A2-3S}
 A_2^{3S} = \beta^{3S}  Z-\frac{\gamma^{3S}  c_{12}^{3S}}{d_{12}^{3S}}\
 =\ \left( Z - 0.008608 \right)\ +\ 0.009021\ \sqrt{Z-0.008608} \ .
\end{equation}
It is evident that the critical charge is defined by $A_1^{3S}$, associated with $(3s)$ Coulomb orbital, this is equal to
\begin{equation}
\label{ZB-3S}
  Z_B^{(3S)}\ =\ 0.928 \ .
\end{equation}

\section{Spin-singlet $4{}^1S$ state}

\subsection{Generalities and results}

The trial function for the spin-singlet $4^1S$ state of the form ($4s'\,1s$) + ($1s\,4s'$) for a Helium-like atom is proposed as a straightforward extension of the trial function (\ref{He3SPsi}),
\begin{equation}
\label{He4SPsi}
\resizebox{0.9\hsize}{!}{$\Psi_{4^1S}\ =\
\frac{1}{2}(1 + P_{12})
\left[ (1 - a_1 Z r_1 + a_2 Z^2 r_1^2 - a_3 Z^3 r_1^3 + a_4 Z r_2 + a_5 Z^2 r_1 r_2 +
b r_{12})\,
e^{-\al Z {\hat r}_1 - \beta Z {\hat r}_2  + \gamma {\hat r}_{12}}\right] \ .$}
\end{equation}
where $P_{12}$ is permutation operator $(1 \lrar 2)$, $a_{1,2,3,4,5 }, b, \al, \beta, \gamma$ are parameters; functions ${\hat r}_{1,2,12}$ are rational functions, see (\ref{rationalrs}).
Preliminary variational calculations indicate that a number of parameters in (\ref{He4SPsi}) can be vanished without effecting the first 4-5 s.d. in the energy: $a_4=a_5=0, b=0$ and \hbox{$ c_2=d_2$}. It leads to the final form of the trial function
\begin{equation}
\label{He4SPsi-final}
\resizebox{0.9\hsize}{!}{$\Psi_{4^1S}\ =\
\frac{1}{2}(1 + P_{12})
\left[ (1 - a_1 Z r_1 + a_2 Z^2 r_1^2 - a_3 Z^3 r_1^3)\,
e^{-\al Z {\hat r}_1 - \beta Z {r}_2  + \gamma {\hat r}_{12}}\right] \ ,$}
\end{equation}
which will be used in the calculations.
The function (\ref{He4SPsi-final}) describes as particular cases the ultra-compact trial functions for the states $2{}^1S (a_2=a_3=0)$ and $3{}^1S (a_3=0)$.  In the limit $Z\to\infty$ it is reduced to
the exact eigenfunction
\begin{equation}
\label{4Sexact}
 \Psi_{4^1S}^ {\rm exact}\ =\ \frac{1}{2}(1 + P_{12})
 \left(  1 - \frac{3}{4}  Z\,r_1  + \frac{1}{8}  Z^2\, r_1^2   -  \frac{1}{192}Z^3\,r_1^3  \right) e^{-\frac{Z}{4} r_1  - Zr_2}\ ,
\end{equation}
leading to the exact energy
\begin{equation}
E_{4^1S} = -\frac{17}{32} Z^2 \, \ .
\end{equation}

The coefficients $a_1, a_2, a_3$ in (\ref{He4SPsi}) are assumed to be determined via imposing the orthogonality conditions: the function $\Psi_{4^1S}$ has to be orthogonal to the functions  $\Psi_{1^1S}$, $\Psi_{2^1S}$, $\Psi_{3^1S}$:
\begin{align}
 \langle \Psi_{4^1S} | \Psi_{1^1S} \rangle & = 0\ ,
\non \\
 \langle \Psi_{4^1S} | \Psi_{2^1S} \rangle & = 0\ ,
\non \\
 \langle \Psi_{4^1S} | \Psi_{3^1S} \rangle & = 0\ ,
\label{ortho4S}
\end{align}
those parameters were found in preliminary calculations, see (\ref{params1Sfits}), (\ref{params2Sfits}), (\ref{params3Sfits}).

The variational results for the energy of the $4{}^1S$ state of Helium-like isoelectronic sequence (for $Z=2,5,10,15,20$) obtained with trial function (\ref{He4SPsi-final}) are presented in Table \ref{Table4S} together with results by APS \cite{APS:1971}, and for Helium $Z=2$,  from DY \cite{Drake:1994,Drake:2006}  and from ABK \cite{Korobov:2018}. Making comparison
the results at $Z=2$ by APS and DY with one by ABK one can see that former accuracies are grossly exaggerated: they are correct only in 5 s.d. and 15 s.d., respectively.
The energies for $Z>10$ are found for the first time. We assume that they are accurate in 3 d.d.

\begin{table}[htp]
\caption{\label{Table4S}
         {\it Variational energy $E$ of the $4{}^1S$ state of He-like
              isoelectronic sequence at $Z=3/2,2,5,10,15,20$ obtained with trial function
              (\ref{He4SPsi-final}).
              Comparison presented with results by APS \cite{APS:1971} rounded to 7 d.d.,
              marked by ${}^\dagger$ at $Z=2,5,10$, with results by DY  \cite{Drake:1994,Drake:2006} marked by~${}^\ddagger$,
              and with benchmark result at $Z=2$ \cite{Korobov:2018}(rounded)
              marked by ${}^\S$ }}
\begin{center}
 {\setlength{\tabcolsep}{0.1cm} \renewcommand{\arraystretch}{1.4}
 \resizebox{0.9\textwidth}{!}{%
\begin{tabular}{|l|l|l|l|l|l|l|} \hline
 Z          &\ ${3}/{2}$ &\ $2$ &\ $5$ &\ $10$ &\ $15$ &\ $20$
\\
\hline\hline
$E$ (a.u.)  &\ -1.133\,877 &\ -2.033\,542 &\ -13.009\,599 &\ -52.553\,312 &\ -118.659\,675
&\ -211.328\,419
\\
            &              &\ -2.033\,575\,8${}^\dagger$ &\ -13.009\,719\,0${}^\dagger$
            &\ -52.553\,511\,8${}^\dagger$ &\ &
\\
            &              &\ -2.033\,586\,717\,030\,726${}^\ddagger$ &                             &                             &               &
\\
            &              &\ -2.033\,586\,717\,030\,725\,447\,439${}^\S$\ &                             &                             &               &
\\[5pt]
\ $a_1$     & \ 0.6061    & \ 0.49926     & \ 0.652916    & \ 0.702231     & \ 0.717785
            & \ 0.725498
\\
$\ a_2$     & \ 0.06719   & \ 0.04911     & \ 0.091236    & \ 0.10769      & \ 0.11325
& \ 0.11607
\\
\ $a_3$     & \ 0.00174   & \ 0.001157    & \ 0.0031434   & \ 0.0041029    & \ 0.0044493
& \ 0.0046291
\\
\ $\al$     & \ 0.60423   & \ 0.5041      & \ 0.34208     & \ 0.28949      & \ 0.26979
& \ 0.26037
\\[5pt]
\ $c_1$     & \ 0.08787   & \ 0.16326     & \ 0.6236      & \ 1.31462      & \ 1.94499
& \ 2.32381
\\[5pt]
\ $d_1$     & \ 0.56486   & \ 0.62715     & \ 1.06429     & \ 1.69731      & \ 2.25439
& \ 2.55319
\\[5pt]
\ $\beta$   & \ 0.99978   & \ 0.99958     & \ 0.99972     & \ 0.99976      & \ 0.99981
& \ 0.99983
\\[5pt]
\ $\gamma$  & \ 0.82000   & \ 0.85159     & \ 0.55639     & \ 0.46381      & \ 0.43458
& \ 0.41955
\\[5pt]
\ $-c_{12}$ & \ 0.0011    & \ 0.00068     & \ 0.01671     & \ 0.06008      & \ 0.08924
& \ 0.10739
\\[5pt]
\ $d_{12}$  & \ 0.88724   & \ 0.95979     & \ 1.09099     & \ 1.56617      & \ 2.35068
& \ 3.2927
\\[5pt]
\hline
\end{tabular}}}
\end{center}
 \end{table}%
After fixing $a_{1,2,3}$ via the orthogonality conditions (\ref{ortho4S}), seven remaining free parameters of the function (\ref{He4SPsi}) can be found variationally.
By making variational calculations for different values of $Z \in [2, 20]$ all parameters can be  fitted via simple functions:
 {\footnotesize
\begin{align}
 \al_{4S}^{\rm fit}\,Z &\ =\ 0.547035 + 0.233319\, Z \ ,
 \non\\
 \beta_{4S}^{\rm fit}\,Z  &\ =\ -0.000706 + 0.999858\, Z \ ,\non
\label{params4Sfits}\\
 \gamma_{4S}^{\rm fit}\,Z &\ =\ 0.935604  + 0.372210\, Z \ ,
\\
 c_{1_{4S}}^{\rm fit} &\ =\ -0.201312 + 0.181028\,Z - 0.002701\,Z^2\ ,
\non\\
 d_{1_{4S}}^{\rm fit} &\ =\ 0.279694 + 0.173856\, Z - 0.002974\,Z^2\ ,
\non\\
 c_{{12}_{4S}}^{\rm fit} &\ =\ 0.020887  - 0.009402\,Z + 0.000146\,Z^2\ ,
\non\\
 d_{{12}_{4S}}^{\rm fit} &\ =\ 0.88804 + 0.018303\,Z + 0.00513\,Z^2 \ ,
\non \\
\non \\
 Z a_{1_{4S}}^{\rm fit}  &\ =\ -0.492327 + 0.750457\,Z\ ,
\non\\
 Z a_{2_{4S}}^{\rm fit}  &\ =\ -0.156545 + 0.123734\,Z\ ,
\non\\
 Z a_{3_{4S}}^{\rm fit}  &\ =\ -0.008765 + 0.005042\,Z\ .
\non
\end{align}
}

\subsection{Majorana formula for $4^1S$ state}

Using the variational energies calculated with trial function (\ref{He4SPsi-final}) with fitted parameters (\ref{params4Sfits}) presented in Table \ref{Emajo3}, one can see that the Majorana formula,
\begin{equation}
\label{Majorana4S}
   E(Z)\ =\ -\frac{17 Z^2}{32}\ +\ 0.060006 Z-0.028533\ ,
\end{equation}
holds for the $4^1S$ state of the Helium-like sequence. The formula (\ref{Majorana4S}) allows us to reproduce 3-4 d.d. in the total energies for $2 \leq Z \leq 20$.

\begin{table}
\caption{\label{Emajo3}
Variational energies $E_{var}\equiv E_{4^1S}$ (a.u.) of the $4^1S$ excited state calculated
for different $Z$ by using the function (\ref{He4SPsi-final}) with parameters (\ref{params4Sfits})  {\it compared}
with energies given by the Majorana formula  (\ref{Majorana4S}).}
\begin{center}
\begin{tabular}{ccc} \hline
 $Z$ & $E_{var}$     & $E_{Majorana}$ (\ref{Majorana4S}) \\
\hline\hline
 2   &  -2.033\,542\,7 &  -2.033\,521  \\
 3   &   ---   &  -4.629\,765  \\
 4   &   ---   &    -8.288\,509  \\
 5   &  -13.009\,599   &  -13.009\,753  \\
 10  &  -52.553\,312 &  -52.553\,473  \\
 15  & -118.659\,675    &  -118.659\,693  \\
 20  &  -211.328\, 419 &  -211.328\,413  \\
\hline
\end{tabular}
\end{center}
\end{table}

\subsection{Square-integrability of the $4{}^1S$ excited state}

The normalizability (square-integrability) of the wave function (\ref{He3SPsi})
is guaranteed if it decays exponentially at large distances $r_1,r_2$. It can be explicitly seen by checking their asymptotic behavior
\[
  \Psi_{{4}^1S} \big\vert_{r_i \ \rm fixed}  \to {\rm exp}\{- A_j  r_j \} \ \ \text{at}
  \ \ r_j\to\infty , \quad  i,j=1,2 \quad (i \neq j)\ ,
\]
cf. (\ref{Adef}),
where $A_{1,2}$ should be simultaneously positive $A_j > 0$.  For the trial function (\ref{He3SPsi})
\begin{align}
\label{Asdefinition-4S}
 A_1 & = \alpha Z \frac{c_1}{d_1} - \gamma \frac{c_{12}}{d_{12}} \equiv A_1^{4S} \ , \\
 A_2 & = \beta  Z \frac{c_2}{d_2} - \gamma \frac{c_{12}}{d_{12}} \equiv A^{4S}_2\ .
\end{align}
These quantities $A_1, A_2$ serve as a measure of the rate of convergence of involved integrals defining the square integrability of $\Psi_{4{}^1S}$. The
trial function $\Psi_{4{}^1S}$  (\ref{He3SPsi}) remains normalizable as long as both
$A_1, A_2$ are positive. Thus, a critical charge $Z=Z_B$ is one at which either $A_1$ or $A_2$
vanishes first. Let us proceed finding the critical charge $Z_B$.

\begin{figure}[hbt]
    \includegraphics[width=0.5\textwidth ,angle=0]{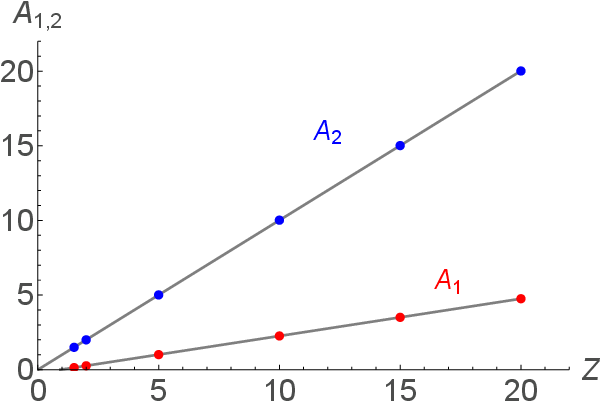}
\caption{\label{A1A2-4S}
          Parameters $A_1^{4S}, A_2^{4S}$ {\it vs.} $Z$, the lines represent
           the interpolations (\ref{A1-4S})-(\ref{A2-4S})}
\end{figure}

By taking the variational parameters for $\Psi_{4{}^1S}$ found for different $Z$, see Table \ref{Table4S}, one can construct the plots of $A$'s {\it vs} $Z$, see Fig. \ref{A1A2-4S}. In general, both $A_1,A_2$ behave essentially as linear functions in $Z$ with a slight deviation from linearity at small $Z$. They can be easily interpolated:
\begin{equation}
\label{A1-4S}
 A_1^{4S} = \frac{\alpha^{4S}  c_1^{4S} Z}{d_1^{4S}}-\frac{\gamma ^{4S} c_{12}^{4S}}{d_{12}^{4S}}\ =\
 \frac{1}{4} (Z-0.928)  -0.0031\sqrt{Z-0.928}\ ,
\end{equation}
\begin{equation}
\label{A2-4S}
 A_2^{4S}\ =\ \beta^{4S}  Z - \frac{\gamma^{4S}  c_{12}^{4S}}{d_{12}^{4S}}\
 =\ (Z-0.002299) + 0.003807\ \sqrt{Z-0.002299} \ ,
\end{equation}
c.f. (\ref{A1-3S})-(\ref{A2-3S}).
It is evident that the critical charge is defined by $A_1^{4S}$, associated the Coulomb orbital $(4s)$,
\begin{equation}
\label{ZB-4S}
   Z_B^{(4S)}\ =\ 0.928\ ,
\end{equation}
cf. (\ref{ZB-3S}).

\section{Spin-singlet $5{}^1S $ state}

After a preliminary analysis the final form of the trial function
for $5{}^1S$ state is chosen as
\begin{equation}
\label{He5SPsi-final}
\resizebox{0.9\hsize}{!}{$\Psi_{5^1S}\ =\
\frac{1}{2}(1 + P_{12})
\left[ (1 - a_1 Z r_1 + a_2 Z^2 r_1^2 - a_3 Z^3 r_1^3 + a_4 Z^4 r_1^4)\,
e^{-\al Z {\hat r}_1 - \beta Z {r}_2  + \gamma {\hat r}_{12}}\right] \ ,$}
\end{equation}
which will be used in further calculations. The function (\ref{He5SPsi-final}) describes as particular cases the ultra-compact trial functions for the states $2{}^1S (a_2=a_3=a_4=0)$, $3{}^1S (a_3=a_4=0)$ and $4{}^1S (a_4=0)$ .  In the limit $Z\to\infty$ it is reduced to
the exact eigenfunction
\begin{equation}
\label{5Sexact}
 \Psi_{5^1S}^ {\rm exact}\ =\ \frac{1}{2}(1 + P_{12})
 \left(  1 - \frac{4}{5}  Z\,r_1  + \frac{4}{25}  Z^2\, r_1^2
 - \frac{4}{375}Z^3\,r_1^3 + \frac{2}{9375}Z^4\,r_1^4 \right)
 e^{-\frac{Z}{5} r_1  - Zr_2}\ ,
\end{equation}
leading to the exact energy
\begin{equation}
E_{5^1S} = -\frac{13}{25} Z^2 \, \ .
\end{equation}

The coefficients $a_1, a_2, a_3, a_4$ in (\ref{He5SPsi-final}) are assumed to be determined via imposing the orthogonality conditions: the function $\Psi_{5^1S}$ has to be orthogonal to the trial functions  $\Psi_{1^1S}$, $\Psi_{2^1S}$, $\Psi_{3^1S}$, $\Psi_{4^1S}$:
\begin{align}
 \langle \Psi_{5^1S} | \Psi_{1^1S} \rangle & = 0\ ,
\non \\
 \langle \Psi_{5^1S} | \Psi_{2^1S} \rangle & = 0\ ,
\non \\
 \langle \Psi_{5^1S} | \Psi_{3^1S} \rangle & = 0\ ,
\non \\
 \langle \Psi_{5^1S} | \Psi_{4^1S} \rangle & = 0\ ,
\label{ortho5S}
\end{align}
which are already found in previous calculations.


\begin{table}[htp]
\caption{\label{Table5S}
         {\it Variational energy $E$ of the $5{}^1S$ state of He-like
              isoelectronic sequence at $Z=3/2,2,5,10,15,20$ obtained with trial function
              (\ref{He5SPsi-final}).
              Comparison presented with results by APS \cite{APS:1971} rounded to 7 d.d.,
              marked by ${}^\dagger$ at $Z=2,5,10$, and with results by DY  \cite{Drake:1994,Drake:2006} at $Z=2$ marked by~${}^\ddagger$.
             }}
\begin{center}
 {\setlength{\tabcolsep}{0.1cm} \renewcommand{\arraystretch}{1.25}
 \resizebox{0.9\textwidth}{!}{%
\begin{tabular}{|l|l|l|l|l|l|l|} \hline
$Z$         &   $\frac{3}{2}$        & $2$                          & $5$                     & $10$                    & $15$        & $20$   \\  \hline\hline
$E$ (a.u.)  &   -1.13377            &  -2.0211552                  & -12.824870              & -51.631224              & -116.437537 & -207.24288  \\
            &                        &  -2.0211632${}^\dagger$      & -12.8249598${}^\dagger$ & -51.6314376${}^\dagger$ &             &   \\
            &                        &  -2.02117685157${}^\ddagger$ &                         &                         &             &    \\[5pt]
 $a_1$      &   0.400000             &   0.522528                   &  0.69394                &  0.746491               &  0.762153   &  0.768826   \\
  $a_2$     &   0.031360              &   0.060889                   &  0.11616                &  0.137189               &  0.143973   &  0.146978   \\
  $a_3$     &   0.000717              &   0.002270                   &  0.00639                &  0.008355               &  0.009038   &  0.009352   \\
  $a_4$     &   4.35$\times 10^{-6}$ &   0.000025                   &  0.00010                &  0.000152               &  0.000170   &  0.000178   \\
  $\al$     &   0.5759               &   0.471894                   &   0.28463               &  0.222209               &  0.201402   &  0.190998  \\[5pt]
   $c_1$    &   0.088109             &   0.156333                   &  0.45361                &  0.73897                &  0.851393   &  0.97418   \\[5pt]
  $d_1$     &   0.642404             &   0.542258                   &  0.8096                 &  0.918099               &  0.921335   &  0.97468   \\[5pt]
  $\beta$   &   0.99948              &   0.999510                   &  0.99985                &  0.999971               &  0.999928   &  0.999942  \\[5pt]
  $\gamma$  &   1.048                &   0.839039                   &  0.4629                 &  0.337522               &  0.29573    &  0.27483   \\[5pt]
  $-c_{12}$ &   0.001872             &   0.001715                   &  0.01512                &  0.04395                &  0.058415   &  0.07      \\[5pt]
  $d_{12}$  &   1.102014            &   1.030673                   &  0.79441                &  0.95531                &  1.64152    &  3.13      \\[5pt]
\hline
\end{tabular}}}
\end{center}
\end{table}%
Using the variational energies calculated with trial function (\ref{He5SPsi-final}) with parameters from Table \ref{Table5S}, one can show that the Majorana formula
of the form,
\begin{equation}
\label{Majorana5S}
   E(Z)\ =\ -\frac{13 Z^2}{25}\  + 0.038738 Z - 0.018601\ ,
\end{equation}
holds for the energy of the $5^1S$ excited state as well, see Table \ref{Emajo5}.

The variational parameters of the $5^1S$ excited state obtained at $Z=2, 5, 10, 15, 20$ can be fitted by the following functions:
\begin{align}
\al_{5S}^{\rm fit}\, Z &\ =\ 0.624211 + 0.159788\,Z\ , \non
\\
\be_{5S}^{\rm fit}\, Z &\ =\ -0.0007046 + 0.999985\,Z\ , \non
\\
\ga_{5S}^{\rm fit}\, Z &\ =\ 1.253784 + 0.212144\,Z\ ,
\label{params5Sfits}\\
c_{1_{5S}}^{\rm fit}\, Z &\ =\ -1.162224 + 0.638270\,Z + 0.019690\,Z^2\ , \non
\\
d_{1_{5S}}^{\rm fit}\, Z &\ =\ -0.387899 + 0.876933\,Z + 0.005675\,Z^2\ , \non
\\
c_{{12}_{5S}}^{\rm fit}\, &\ =\ 0.017577 - 0.007836\,Z + 0.000183\,Z^2\ , \non
\\
d_{{12}_{5S}}^{\rm fit}\, &\ =\ 1.276337 - 0.150483\,Z + 0.011677\,Z^2\ , \non
\\
a_{1_{5S}}^{\rm fit}\, Z &\ =\ -0.553144 + 0.798128\,Z\ , \non
\\   
a_{2_{5S}}^{\rm fit}\, Z &\ =\ -0.193869 + 0.156684\,Z\ , \non
\\
a_{3_{5S}}^{\rm fit}\, Z &\ =\ -0.016236 + 0.010114\,Z\ , \non
\\
a_{4_{5S}}^{\rm fit}\, Z &\ =\ -0.000363 + 0.000194\,Z\ , \non
\end{align}

\begin{table}
\caption{\label{Emajo5} Variational energies $E_{var}\equiv E_{5^1S}$ (a.u.) of the $5^1S$ state calculated for different $Z$
based on the function (\ref{He5SPsi-final}) with variational parameters  {\it compared} with ones of the Majorana
formula  (\ref{Majorana5S}).   }
\begin{center}
\begin{tabular}{c|c|c} \hline
 $Z$ & $E_{var}$ & $E_{Majorana}$ (\ref{Majorana5S}) \\
\hline\hline
 $\frac{3}{2}$  \ &\ -1.13377 \ &\ -1.13049 \\
 2 \ &\ -2.02116\ &\ -2.02113 \\
 5 \ &\ -12.8249\ &\ -12.8249 \\
 10\ &\ -51.6312\ &\ -51.6312 \\
 15\ &\ -116.438\ &\ -116.438 \\
 20\ &\ -207.243\ &\ -207.244 \\
\hline
\end{tabular}
\end{center}
\end{table}

\begin{figure}[hbt]
    \includegraphics[width=0.5\textwidth ,angle=0]{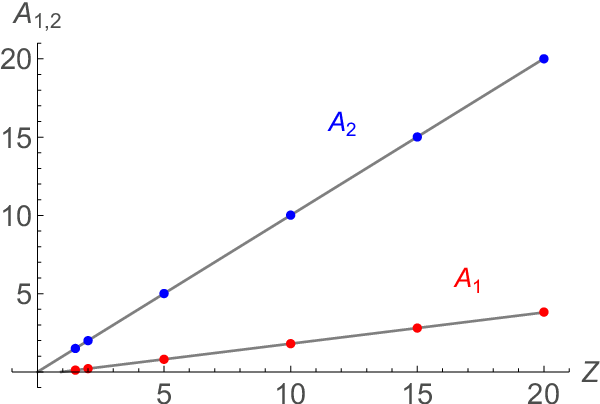}
\caption{\label{A1A2f5S}
    Parameters $A_1^{5S}, A_2^{5S}$ for different $Z$ and their interpolations (\ref{A1-5S})-(\ref{A2-5S})}
\end{figure}

By taking the variational parameters in $\Psi_{5^1S}$ (\ref{He5SPsi-final}) found
for different $Z$, see Table \ref{Table5S}, one can construct
the plots of $A_{1,2}$ {\it vs.} $Z$, see Fig. \ref{A1A2f5S}. In general,
both $A_1,A_2$ behave as linear functions
in $Z$ with a slight deviation from linearity at small $Z$. They can be easily
interpolated:
\begin{equation}
\label{A1-5S}
  A_1^{5S}\ =\ \frac{\al^{5S}\,c_1^{5S}\,Z}{d_1^{5S}}\ -\ \frac{\gamma^{5S}\, c_{12}^{5S}}{d_{12}^{5S}}\
  =\ \frac{1}{5}\,\left(Z\,-\,0.9387\right) \ -\ 0.002730\,\sqrt{Z\,-\,0.9387}\ ,
  \end{equation}
\begin{equation}
\label{A2-5S}
 A_2^{5S}\ =\ \beta^{5S}\,Z\ -\ \frac{\gamma^{5S}\,c_{12}^{5S}}{d_{12}^{5S}}\
 =\ \left(Z\,-\,0.00655 \right)\ +\ 0.005297\,\sqrt{Z\,-\,0.00655} \ .
\end{equation}
It is evident that the critical charge is defined by $A_1^{5S}$, associated with $(5s)$ Coulomb orbital, this is equal to
\begin{equation}
\label{ZB-5S}
  Z_B^{(5S)}\ =\ 0.939\ .
\end{equation}

\section{Spin-singlet $n{}^1S $ state}

Based on the trial functions presented for $1^1S$ (\ref{He1SPsi-final}), $2^1S$ (\ref{He1SPsi-2S}), $3^1S$ (\ref{He3SPsi-final}), $4^1S$ (\ref{He4SPsi-final}), $5^1S$ (\ref{He5SPsi-final}) states with parameters (\ref{params1Sfits}), (\ref{params2Sfits}), (\ref{params3Sfits}), (\ref{params4Sfits}), (\ref{params5Sfits}), respectively, we conjecture that the trial function for $n^1S$ state with $n=1,2, \ldots $, which gives the accuracy in energy at least 3 d.d. (non-rounded), should be of the form
\[
\Psi_{n^1S}\ =\
\frac{1}{2}(1 + P_{12})
\]
\begin{equation}
\label{HenSPsi}
 \bigg( (1 - a_1  Z r_1 + a_2 Z^ 2 r_1^2 + \ldots + (-)^{n-1}a_{n-1} Z^{n-1} r_1^{n-1} )\,
e^{-\al Z {\hat r}_1 - \beta Z {\hat r}_2  + \gamma {\hat r}_{12}}\bigg) \ ,
\end{equation}
where $P_{12}$ is the permutation operator $(1 \lrar 2)$, $a_{1,2,\ldots, n-1}, \al, \beta, \gamma$ are parameters. Parameters $a_{1,2,\ldots, n-1}$ in (\ref{HenSPsi}) are non variational: they are fixed by imposing the orthogonality conditions to all previously found $(n-1)$ excited states $1^1S$, $2^1S$, $\ldots$, ${(n-1)}^1S$. The effective distances
\begin{equation*}
  {\hat r}_i = \frac{1+c_i r_i}{1+ d_i r_i}\ r_i\ , \ i=1,2,12\ ,
\end{equation*}
which enter to (\ref{HenSPsi}), are characterized by the three pairs of parameters $(c_i,d_i)$, where $i={1,2,12}$.

From physics point of view the proposed function (\ref{HenSPsi}) is of the type ($ns'\,1s$) + ($1s\,ns'$) multiplied by the Jastrow factor $J(r_{12})$,
\[
 \Psi_{n^1S}\ =\ \left[(ns'\,1s) + (1s\,ns')\right]\, J(r_{12})\ ,
\]
where
\[
  J(r_{12})\ =\ e^{\gamma {\hat r}_{12}}\ .
\]
For all states, even for the ground state, where the trial function is of the type
\[
   [(1s\,1s')+(1s'\,1s)]\, J(r_{12})  \ ,
\]
one electron is in $(1s)$ Coulomb state in $r_2$ coordinate, thus, it is (much) closer
to the nuclei than another one, hence, the screening for this electron in interaction with nuclei can be neglected:
${\hat r_2} \rar r_2$, $c_2 \approx d_2$ and $\beta \approx 1$.
It is confirmed in the calculations of five $1^1S$ (\ref{He1SPsi-final}), $2^1S$ (\ref{He1SPsi-2S}), $3^1S$ (\ref{He3SPsi-final}), $4^1S$ (\ref{He4SPsi-final}), $5^1S$ (\ref{He5SPsi-final}) states.
Another electron is further away from the nuclei and its interaction with nuclei is screened by the presence of the first electron, thus, $c_1 \neq d_1$.
For the ground state $1^1S$ both electrons are formally in $(1s)$ state but due to the phenomenon of {\it clusterization}, which occurs owing to inter-electron interaction (and the effective Pauli repulsion), one electron is closer to the nuclei than another one. Thus, a ``non-linear", distance-dependent screening could play some role, it is assumed that $c_1 \neq d_1$, see \cite{Part-1:2020}. However, it was checked by making a concrete calculation that the first 3 d.d. in the ground energy are not influenced by presence/absence of a ``non-linear" screening for any $Z<20$ and $c_1 = d_1$. From another side in order to reach
the accuracy in energy of 3 d.d. we need to introduce a "weak" screening by adding to the prefactor the $r_1(r_2)$ and $r_{12}$ terms,
\[
     1\ \rar (1 + a r_1 + b r_{12}) \ ,
\]
see (\ref{He1SPsi-final}).

In the limit $Z \rar \infty$, the original 3-body problem $(Z,2e)$ is reduced to two non-interacting $Z$-Hydrogen atoms $(Z,e)$ and can be solved exactly. At this limit in the function (\ref{HenSPsi}) it should happen that the parameter $\gamma \rar 0$, the non-linear screening disappears ${\hat r}_{1,12} \rar r_{1,12}$, the Jastrow factor $J(r_{12}) \rar 1$ and the polynomial prefactor becomes the associated Laguerre polynomial. Eventually, the function $\Psi_{n^1S}$ (\ref{HenSPsi}) is reduced to the exact eigenfunction,
\[
  \Psi_{n^1S}^{(exact)}\ =\ (ns\,1s) + (1s\,ns)\ =\ \frac{1}{2}(1 + P_{12})\
  \bigg[L_{n-1}^{(1)}\bigg(\frac{2 Z r_1}{n} \bigg) \,e^{-\frac{Zr_1}{n} - Z r_2 }\bigg]\ ,
\]
where $L_{n-1}^{(1)}(2 Z r_1/n )$ is the associated Laguerre polynomial,
and the energy becomes
\[
  E_{n^1S}^{(exact)}(Z)\ =\
  -{\frac {Z^{2}}{2}} \left(1\ +\ \frac{1}{n^2} \right)\ .
\]
It is evident that the presence of the Coulomb repulsion of the electrons at finite $Z$
leads to the inequality,
\[
    E_{n^1S}(Z)\ \geq \ E_{n^1S}^{(exact)}(Z)\ .
\]


\section{Majorana Formulae}

\begin{figure}[hbt]
    \includegraphics[width=0.5\textwidth ,angle=0]{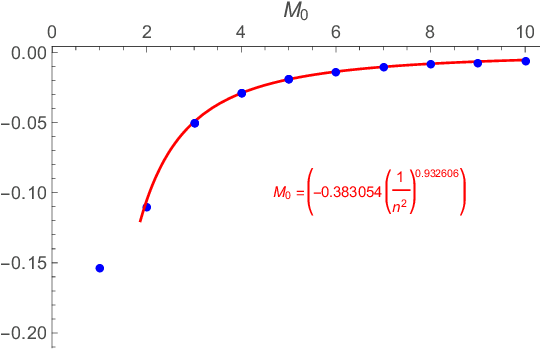}
\caption{\label{MajoM0fig}
     Coefficient $M_0(n)$ in the Majorana formula (\ref{Majon}) {\it vs.} $n$ }
\end{figure}

\begin{figure}[hbt]
    \includegraphics[width=0.5\textwidth ,angle=0]{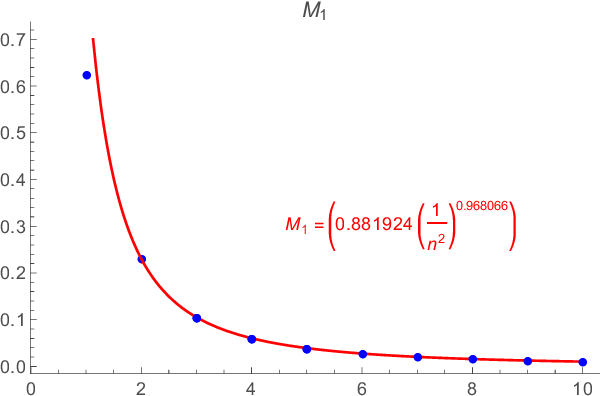}
\caption{\label{MajoM1fig}
     Coefficient $M_1(n)$ in the Majorana formula (\ref{Majon}) {\it vs.} $n$ }
\end{figure}

As was stated in \cite{AOP:2019} for the ground state energy $E_{1{}^1S }(Z)$ the Majorana formula holds reproducing 4-5 s.d. The same is correct for the energy $E_{2{}^1S }(Z)$ \cite{Part-2:2021}. We checked in this article, see Sections II-IV, that the energy
$E_{n{}^1S }(Z)$ of the $n{}^1S $ excited states ($n=3, 4, 5$) of the Helium-like sequence
at $Z \leq 20$ can be fitted with high accuracy by the Majorana formula (a second degree polynomial in $Z$) as well:
\begin{equation}
\label{Majon}
E^{\rm \small Majorana}_{n{}^1S }(Z)\ =\ -M_0(n) + M_1(n)  Z - M_2(n) Z^2
\end{equation}
see the formulae (\ref{Majorana3S}), (\ref{Majorana4S}), (\ref{Majorana5S}) and Tables IV, VI, VIII, respectively. The coefficients $M_0(n), M_1(n)$ and $M_2(n)$ are collected in Table \ref{Majorana3nS}. Let us emphasize that these coefficients are obtained by making fit (\ref{Majon}) of the most accurate variational energies available in the literature.
\begin{table}
\begin{center}
\begin{tabular}{c|c|c|c}
\hline
 ${n}$ & $M_0$ \ &\ $M_1$ & $\ M_2$ \ \\
  \hline
  \rule{0pt}{1.2\normalbaselineskip}
 1 & 0.153282\ & 0.624583 & $1$
\\[4pt]
 2 & 0.109971\ & 0.231405 & $\frac{5}{8}$
\\[4pt]
 3 & 0.049610\ & 0.105296 & $\frac{5}{9}$
\\[4pt]
 4 & 0.028533\ & 0.060006 & $\frac{17}{32}$
\\[4pt]
 5 & 0.018553\ & 0.038714 & $\frac{13}{25}$
\\[4pt]
\hline
\end{tabular}
\end{center}
\caption{\label{Majorana3nS}
     $M_{0, 1, 2} (n)$ coefficients in Majorana formula (\ref{Majon}) for $n=1,2,3,4,5$. }
\end{table}
Interestingly, for $n\geq 2$ these coefficients $M_{0,1,2}$ themselves can be
fitted by using the formulas:
\begin{align}
 M_0(n) &\ =\ 0.383054\left (\frac{1}{n^2} \right)^{0.932606}\ , \non
\\
 M_1(n) &\ =\ 0.881924 \left(\frac{1}{n^2}\right)^{0.968066}\ ,
\label{Majocofitted}\\
 M_2(n) &\ =\ \frac{1}{2}\,  \left(1 + \frac{1}{n^2}\right)\ ,
\non
\end{align}
see e.g. Figs. \ref{MajoM0fig}, \ref{MajoM1fig} for $M_{0,1}$, respectively.
Quality of fit (\ref{Majocofitted}) is illustrated in Table \ref{HevsMajo}: systematically,
the deviation from the exact energies occurs in the 4th d.d.


\begin{table}
\begin{center}
 {\setlength{\tabcolsep}{0.1cm} \renewcommand{\arraystretch}{1.3}
 \resizebox{1.0\textwidth}{!}{%
\begin{tabular}{|c|cc|c|c|c|c|} \hline
  $n$ & $E (Z=2)$  \cite{Drake:1994}  &  $E^{\rm \tiny Majorana}(Z=2)$
  &  $E^{\rm \tiny Majorana}(Z=5)$
  &  $E^{\rm \tiny Majorana}(Z=10)$
  &  $E^{\rm \tiny Majorana}(Z=15)$
  &  $E^{\rm \tiny Majorana}(Z=20)$ \\ \hline
2   & -2.14597404605  &  -2.1442 & -14.5778 & -60.3005 & -137.273 & -245.496  \\
3   & -2.06127198974  &  -2.0614 & -13.4127 & -54.5538 & -123.473 & -220.169  \\
4   & -2.03358671703  &  -2.0334 & -13.0090 & -52.5516 & -118.657 & -211.324  \\
5   & -2.02117685157  &  -2.0208 & -12.8236 & -51.6281 & -116.433 & -207.237  \\
6   & -2.01456309845  &  -2.0142 & -12.7234 & -51.1278 & -115.227 & -205.020  \\
7   & -2.01062577621  &  -2.0102 & -12.6634 & -50.8268 & -114.500 & -203.684  \\
8   & -2.00809362211  &  -2.0077 & -12.6245 & -50.6318 & -114.030 & -202.818  \\
9   & -2.00636955311  &  -2.0060 & -12.5980 & -50.4984 & -113.707 & -202.225  \\
10  & -2.00514299175  &  -2.0048 & -12.5791 & -50.4031 & -113.477 & -201.801  \\
\hline
\end{tabular}}}
\end{center}
\caption{
\label{HevsMajo}
     The energies $E$ (a.u.) of the spin-singlet $n{}^1S$ excited states of Helium atom at $n=2,\ldots,10$ taken from \cite{Drake:1994} at $Z=2$ {\it vs.} the prediction
     by the Majorana formula (\ref{Majon}) with coefficients (\ref{Majocofitted}) at $Z=2,5,10,15,20$.}
\end{table}

In Tables \ref{MajoGS}-\ref{Majo2S} the quality of description of the energies of the $1{}^1S$ and $2{}^1S$ states for different nuclear charges $Z \in [2-10]$ using the Majorana formula (\ref{Majon}) is demonstrated:

\begin{table}
\begin{center}
\begin{tabular}{c|c|c}
\hline
 $Z$ & $E$ \cite{APS:1971} & $E_{Majorana}$
\\
\hline \hline
  \rule{0pt}{1.2\normalbaselineskip}
 2 &\ -2.90372 \ & -2.90412
\\[0.5pt]
 3 & \ -7.27991 \ & -7.27953
\\[0.5pt]
 4 &\ -13.6556 \ & -13.6550
\\[0.5pt]
 5 &\ -22.031 \  & -22.0304
\\[0.5pt]
 6 &\ -32.4062 \ & -32.4058
\\[0.5pt]
 7 &\ -44.7814 \ & -44.7812
\\[0.5pt]
 8 &\ -59.1566 \ & -59.1566
\\[0.5pt]
 9 &\ -75.5317 \ & -75.5320
\\[0.5pt]
 10 &\ -93.9068 \ & -93.9075
\\
\hline
\end{tabular}
\end{center}
\caption{\label{MajoGS}
$1{}^1S$ ground state energy $E$ (a.u.) \cite{APS:1971} for the He-like sequence {\it vs.} prediction of the Majorana formula
$E^{\rm \small Majorana}_{1{}^1S }(Z) = - 0.153282 + 0.624583\,Z - Z^2$
(see Annals of Physics (2019) \cite{AOP:2019}).}
\end{table}

\begin{table}
\begin{center}
\begin{tabular}{l|l|l} \hline
 $Z$ & \quad $E$  \cite{APS:1971} &\ $E_{Majorana}$
\\
 \hline\hline
\rule{0pt}{1.2\normalbaselineskip}
 2     &\ -2.14598  \  & -2.14422
\\[0.5pt]
 3     &\ -5.04088  \  & -5.03876
\\[0.5pt]
 4     &\ -9.18488  \  & -9.18330
\\[0.5pt]
 5     &\ -14.57853 \  & -14.57784
\\[0.5pt]
 6     &\ -21.22202 \  & -21.22238
\\[0.5pt]
 7     &\ -29.11542 \  & -29.11691
\\[0.5pt]
 8     &\ -38.25876 \  & -38.26145
\\[0.5pt]
 9     &\ -48.65206 \  & -48.65599
\\[0.5pt]
 10    &\ -60.29534 \  & -60.30053
\\
\hline
\end{tabular}
\end{center}
\caption{\label{Majo2S}
 $2{}^1S$ state energy $E$ (a.u.) for He-like sequence: results from \cite{APS:1971}
{\it vs.} prediction by Majorana formula (\ref{Majon}) with coefficients
(\ref{Majocofitted}) at $n=2$.}
\end{table}

\pagebreak

The coefficients (\ref{Majocofitted}) in the Majorana formula (\ref{Majon}) can be used to predict the energies of $n{}^1S $ excited states for $n > 5$ for different $Z > 2$. For example, in Table X the Majorana energies for $Z=5,10, 15, 20$ are shown.

\vfil


\section*{Conclusions}

In this article we constructed, based on physics arguments, an ultra-compact, seven-parametric, permutationally-symmetric wavefunctions for the spin-singlet $S$ states for two-electron sequence with $Z \leq 20$:

\begin{itemize}
\item ${1^1S}$ state,

\begin{equation}
\label{psi1}
\Psi_{1}^{(+)}\ =\
\frac{1}{2}(1 + P_{12})
\bigg[ (1 - a_1  Z r_1 +  b_1 r_{12})\,
e^{-\al_1 Z {r}_1 - \beta_1 Z {r}_2  + \gamma_1 {\hat r}_{12}^{(1)}} \bigg] \ ,
\end{equation}
where $a_1, b_1$ are variational parameters, $\al_1, \beta_1, \gamma_1$ are variational parameters.

\item ${n^1S}$ states, $n > 1$,

\begin{equation}
\label{psin}
\Psi_{n}^{(+)}\ =\ \frac{1}{2}(1 + P_{12})
 \bigg[ P^{(n)}_{n-1}\bigg(\frac{2 Z r_1}{n}\bigg)
e^{-\al_n Z {\hat r}^{(n)}_1 - \beta_n Z {r}_2  + \gamma_n {\hat r}^{(n)}_{12}}\bigg] \ ,
\end{equation}

where $P_{12}$ is the permutation operator $(1 \lrar 2)$, $\al_n, \beta_n, \gamma_n$ are variational parameters. For given $n$ the polynomial $P^{(n)}_{n-1}$ of degree $(n-1)$ is fixed uniquely by imposing the orthogonality conditions to all previously found $(n-1)$ excited states $1^1S$, $2^1S$, $\ldots$, ${(n-1)}^1S$. The effective distances are given by
\begin{equation*}
  {\hat r}^{(n)}_1 = \frac{1+c^{(n)}_1 r_1}{1+ d^{(n)}_1 r_1}\ r_1\quad , \quad
  {\hat r}^{(n)}_{12} = \frac{1+c^{(n)}_{12} r_{12}}{1+ d^{(n)}_{12} r_{12}}\ r_{12}\ , \
\end{equation*}
where $c^{(n)}_1, d^{(n)}_1$ and $c^{(n)}_{12}, d^{(n)}_{12}$ are variational parameters.

\end{itemize}

Despite the extraordinary simplicity of the trial functions (\ref{psi1})-(\ref{psin}) they provide an accuracy of 3 d.d. in the energy for all studied spin-singlet $S$ states at various values
of $Z \leq 20$ in comparison with existing benchmark variational calculations. It definitely makes sense to use the Lagrange Mesh Method to verify these calculations as well as to carry out the calculation of the energies of not-yet-studied states in a similar way as was done for the $1^1S$ state in \cite{OT:2015}. This will be done elsewhere.
This implies that the functions (\ref{psi1})-(\ref{psin}) reproduce the domain of applicability of non-relativistic quantum mechanics of point-like Coulomb charges QMCC with static (infinitely-massive) nuclei, where mass, relativistic and QED corrections do not contribute. Surprisingly, the variational parameters $\al_n Z\,,\, \beta_n Z$ demonstrate linear in $Z$ behavior for $n=1,2,3,4,5$ while the parameter $\gamma_n Z$ is a linear function in $Z$ for $n=2,3,4,5$. The quality of (\ref{psi1})-(\ref{psin}) is reflected in the closeness of the parameter $\beta_n\,Z$ to the exact electron-nuclear cusp parameter (equal to $Z$) and of the expression $\gamma_n c^{(n)}_{12}/d^{(n)}_{12}$ to the exact electron-electron cusp parameter (equal to $1/2$), for the definitions of the cusps see Part I \cite{Part-1:2020}.

Further complexification of the trial functions (\ref{psi1})-(\ref{psin}) by adding extra terms into the pre-exponential factor (prefactor) and/or modifying terms in the exponent does not make much sense: it leaves 3 d.d. in energy unchanged, while the changes appear in the 4th d.d. (and the subsequent ones), which are subject to change due to various corrections to QMCC anyway. However, it definitely makes sense to consider the Hamiltonian (1) with the addition of the nuclear kinetic energy
\begin{equation}
\label{H-He-complete}
  {\cal H}\ =\ -\frac{1}{2M}\boldsymbol{\nabla}_R^2 - \frac{1}{2} \sum_{i=1}^{2} \boldsymbol{\nabla}_i^2 \ -\ \frac{Z}{r_1} - \frac{Z}{r_2} \ +\ \frac{1}{r_{12}}  \ ,
  \end{equation}
where $M$ is the nuclear mass, variationally to take into account finite-mass effects.
In order to do so the center-of-mass should be separated out and
the generalized Euler coordinates should introduced in space of relative motion, see e.g. \cite{TMA:2016}, the angular dependence of the wave functions should be excluded~
\footnote{It corresponds to the case of $S$ states} and $r_1, r_2, r_{12}$ play the role of relative distances. As the result we arrive at three-dimensional Hamiltonian in the $r_1, r_2, r_{12}$ variables with a non-standard kinetic energy, c.f.(1). The {\it same} trial functions (\ref{psi1})-(\ref{psin}) can then be used in variational calculations. Evidently, they will ascribe a mass dependence to the coefficients $\al_n, \beta_n, \gamma_n$, and $c^{(n)}_1, d^{(n)}_1$ and $c^{(n)}_{12}, d^{(n)}_{12}$. Alternatively, the corresponding Schr\"odinger equation in the relative coordinates can be solved by using the Lagrange Mesh method \cite{OT:2015} with high accuracy in order to verify the variational calculations. This will be realized elsewhere.

For the ${n^1S}$ states with $n = 3,4,5$ the Second Critical charge $Z_B^{(n)}$, where the ultra-compact trial functions loose their square-integrability, was calculated: $Z_B^{(1)}=Z_B^{(2)}\,=\,0.9049$, $Z_B^{(3)}=Z_B^{(4)}\,=\,0.928$, $Z_B^{(5)}\ =\ 0.939$.
In \cite{TLO-analytic:2016} it was shown that the coinciding Second Critical charges
$Z_B^{(1)}=Z_B^{(2)}$ imply the appearance of a square-root branch point at $Z=Z_B^{(1)}$ of the energy $E_{1{}^1S }(Z)$ in the complex plane of $Z$. In this point the energy levels
$E_{1{}^1S }(Z)$ and $E_{2{}^1S }(Z)$ intersect, the level crossing occurs and the Puiseux expansion (the expansion in half-integer degrees) can be constructed,
\begin{equation}
\label{Puiseux}
   E(Z)=E_B+p_1(Z-Z_B)+q_3(Z-Z_B)^{3/2} + p_2 (Z-Z_B)^2 + q_5 (Z-Z_B)^{5/2}\ +\ \ldots
\end{equation}
where
\[
 Z_B^{(1,2)}\ =\ {0.9049}\ ,\ E_B^{(1,2)}\ =\ {-0.4079} \ ,\
 p_1^{(1,2)}\ =\, {-1.1235} \ ,
 \]
\[
 \ q_3^{(1)}\ =\ -0.1978\ ,\ p_2^{(1,2)}\ =\ -0.7528\ ,\ q_5^{(1)}\ =\ -0.1083\ ,
\]
cf.\cite{AOP:2019}, where $q_3^{(2)}\ =\ -q_3^{(1)}$ and $q_5^{(2)}\ =\ -q_5^{(1)}$.
It was also shown that $1/Z_B^{(1)}$ defines the radius of convergence of $1/Z$-expansion for the ${n^1S}, n=1,2$ states. Furthermore, at $Z=Z_B^{(1)}=Z_B^{(2)}$ the functions $\Psi_{1}^{(+)}$, $\Psi_{2}^{(+)}$, see (\ref{psi1})-(\ref{psin}), become degenerate. Do exist other branch points
of $E_{1{}^1S }(Z)$ Riemann sheet than at $Z=Z_B^{(1)}$ remains an open question. It is evident from the complex analysis point of view that the energies $E_{2{}^1S }(Z)$ and $E_{3{}^1S }(Z)$ should intersect, there must exist pair(s) of complex-conjugated square-root branch points:
where are they situated? This is another open question.
It seems quite suggestive to make a similar analysis for ${n^1S}, n=3,4$ as the one made for ${n^1S}, n=1,2$: does the coincidence of $Z_B^{(3)}=Z_B^{(4)}$ imply a level crossing and square-root branch point of the energy in the complex plane of $Z$? All these questions
might be addressed in future studies of the analytical structure
of the energy in $Z$.

The validity of the Majorana formula,
\[
    E_n(Z) \ =\ -M_0(n)\ +\ M_1(n)\,Z\ -\ M_2(n)\,Z^2
\]
which was established for the ${1^1S}$ and ${2^1S}$ states in Parts I, II, respectively, was extended to the ${n^1S}$ states, at $n = 3,4,5$. For higher $n=6, 7, 8, 9, 10$ it was checked that
the Majorana formula leads to accurate results for $Z=2$. Undoubtedly, a high accuracy in the energies will be obtained for larger $Z > 2$.

The functions (\ref{psi1})-(\ref{psin}) allow us to make a natural modification for the case of spin-triplet $S$ states by making them permutationally-anti-symmetric:

\begin{itemize}
\item ${1^3S}$ state,

\begin{equation}
\label{psi1-}
\Psi_{1}^{(-)}\ =\ \frac{1}{2}(1 - P_{12})
\bigg[ (1 + a_1  Z r_1 +  b_1 r_{12})\,
e^{-\al_1^{(-)}\, Z {r}_1 - \beta_1^{(-)}\, Z {r}_2  + \gamma_1^{(-)}\, {\hat r}_{12}^{(1)}} \bigg] \ ,
\end{equation}
where $a_1, b_1$ are variational parameters, $\al_1^{(-)}, \beta_1^{(-)}, \gamma_1^{(-)}$ are variational parameters, c.f. Part II \cite{Part-2:2021}.

\item ${n^3S}$ states, $n > 1$,

\begin{equation}
\label{psin-}
\Psi_{n}^{(-)}\ =\ \frac{1}{2}(1 - P_{12})
 \bigg[ P^{(n,-)}_{n-1}\bigg(\frac{2 Z r_1}{n}\bigg)
e^{-\al_n^{(-)}\, Z {\hat r}^{(n)}_1 - \beta_n^{(-)}\, Z {r}_2  + \gamma_n^{(-)}\, {\hat r}^{(n)}_{12}}\bigg] \ ,
\end{equation}
where $P_{12}$ is the permutation operator $(1 \lrar 2)$, $\al_n^{(-)}, \beta_n^{(-)}, \gamma_n^{(-)}$ are variational parameters. For given $n$ the polynomial $P^{(n,-)}_{n-1}$ of degree $(n-1)$ is fixed uniquely by imposing the orthogonality conditions to all previously found $(n-1)$ excited states $1^3S$, $2^3S$, $\ldots$, ${(n-1)}^3S$.

\end{itemize}
The correctness of such a modification will be checked elsewhere.


\section*{Acknowledgments}

\noindent
This work is partially supported by CONACyT grant A1-S-17364 and DGAPA grant IN113022 (Mexico). A.V.T. thanks PASPA-UNAM for a support during his sabbatical stay in 2021-2022 at the University of Miami, where this work was initiated. The authors thank J C del Valle for interest in the work and an important comment.

This article is dedicated to the memory of Professor Frank E Harris who passed away on March 9, 2023. His dedication to science was exemplary, he was always open to new knowledge, new ideas,
it was always a pleasure to discuss science with him. We will definitely miss his presence. R.I.P.

\end{document}